\documentclass[twocolumn,showpacs,preprintnumbers,amsmath,amssymb]{revtex4}

\usepackage{graphicx}
\usepackage{dcolumn}
\usepackage{bm}

\begin{document}
\newcommand{\gdhi}{\ooalign{\hfil/\hfil\crcr$\partial$}}

\def\Sp{\mathop{\mathrm{Sp}}\nolimits}
\def\sgn{\mathop{\mathrm{sgn}}\nolimits}
\def\erfc{\mathop{\mathrm{erfc}}\nolimits}
\def\tr{\mathop{\mathrm{tr}}\nolimits}
\def\as{\mathop{\mathrm{as}}\nolimits}
\def\val{\mathop{\mathrm{val}}\nolimits}

\title{Nuclear matter in the chiral quark soliton model with vector mesons}

\author{S.Nagai$^{1}$}
\email{j6206701@ed.noda.tus.ac.jp}
\author{N.Sawado$^{1}$}
\email{sawado@ph.noda.tus.ac.jp}
\author{N.Shiiki$^{1},^{2}$}
\email{norikoshiiki@mail.goo.ne.jp}
\affiliation{$^{1}$Department of Physics, Faculty of Science and Technology, 
Tokyo University of Science, Noda, Chiba 278-8510, Japan\\
$^{2}$Department of Management, Atomi University, Niiza, Saitama 
352-8501, Japan
}
\date{\today}

\begin{abstract}
We study the nuclear matter solution in the chiral quark soliton model 
coupled to $\rho$ and $\omega$ vector mesons based on the Wigner-Seitz 
approximation. It is shown that the vector mesons stabilize the soliton 
at high-density region. As a result, the saturation property and 
incompressibility are significantly improved. 
\end{abstract}


\maketitle

\section{\label{sec:level1}Introduction\protect\\ } 
The idea of investigating dense nuclear matter in the topological 
soliton models has been developed over decades. 
It was first applied for the nuclear matter system with 
the skyrmion centered cubic (CC) crystal by Klebanov~\cite{klebanov85}. 
This configuration was studied further by W\"ust, Brown and Jackson 
to estimate the baryon density and discuss the phase transition 
between nuclear matter and quark matter~\cite{wust87}.  
Goldhabor and Manton found a new configuration, body-centered cubic (BCC)   
of half-skyrmions in a higher density regime~\cite{manton87}. 
The face centered cubic (FCC) and BCC lattice were studied by Castillejo 
{\it et al.}~\cite{castillejo89} 
and the phase transitions between those configurations were 
investigated by Kugler and Shtrikman~\cite{kugler89}. 
Recently, the idea of using crystallized skyrmions to study 
nuclear matter was revived by Park, Min, Rho and Vento 
with the introduction of the Atiyah-Manton multi-soliton ansatz 
in a unit cell~\cite{park02}.   

The soliton model incorporating quark degrees of freedom into each soliton  
was also considered in 80's.  
Achtzehnter, Scheid and Wilets investigated the Friedberg-Lee 
soliton bag model with a simple cubic lattice~\cite{achtzehnter85}. 
Due to the periodicity of the background potential, the solution of 
the Dirac equation has the form of the Bloch waves, 
$\psi_{\bm k}({\bm r})=e^{i\bm{k}\cdot\bm{r}}\phi_{\bm k}({\bm r})$ 
where $\phi_{\bm k}$ satisfies the same periodic boundary condition 
as the background potential.  

The Wigner-Seitz approximation was used for the analysis of the 
crystal soliton model with quarks. In this ansatz, a single soliton is placed 
on the center of a spherical unit cell. 
Then the lowest energy level (``bottom'' of the band) for the valence quarks becomes 
s-state. The appropriate boundary conditions at the cell 
boundary should be imposed on the quark wave functions 
as well as the chiral fields. This simple treatment sheds 
light on the nucleon structure in nuclear medium. Soliton 
matter within this approximation have been extensively 
studied by using various nucleon models such as the 
the chiral quark-meson type model
~\cite{banerjee85,glendenning86,hahn87,weber98}, 
Friedberg-Lee soliton bag model~\cite{reinhardt85,weber98,birse88,barnea00}, 
the Skyrme model~\cite{Kutschera84}.
The non-zero dispersion of the lowest band \cite{weber98} 
and the quark-meson coupling \cite{barnea00} were also examined within this 
approximation. 

The chiral quark soliton model (CQSM) can be interpreted as the soliton bag model 
including not only valence quarks but also the vacuum sea quark polarization 
effects explicitly~\cite{diakonov88,reinhardt88,meissner89,wakamatsu91}. The model provides 
correct observables of a nucleon such as mass, electromagnetic
value, spin carried by quarks, parton distributions
and octet, decuplet $SU(3)$ baryon spectra~\cite{christov96,alkofer96}. 
Amore and De Pace studied nuclear matter in the CQSM using the Wigner-Seitz 
approximation and observed the nuclear saturation~\cite{amore00}. 
They examined the soliton solutions with three different boundary  
conditions imposed on the quark wave function. However the obtained 
saturation density was lower than the experimental value. They thus  
concluded that such discrepancy is originated in the approximate    
treatment~\cite{adjali92} of the sea quark contribution.

In Ref.\cite{nagai06}, we studied the nuclear matter in CQSM and 
observed splitting of the nucleon-$\Delta$ spectra. 
The vacuum polarization was treated exactly and a relatively shallow saturation 
was obtained. 
However for the value of the constituent quark mass $M$ reproducing the octet 
and decuplet baryon spectra, the soliton breaks even at low densities.

In this paper, we construct the matter soliton solutions including $\pi, \sigma, \rho$ and $\omega$.
The role of $\rho, \omega$ is to produce the short range effects of the nuclear force 
and stabilize the solution at high densities.
It is straightforward to include the $\rho$ meson in the CQSM,  
but the $\omega$ meson requires some technique since the Hamiltonian is no longer 
real. To overcome this difficulty, we apply two different methods proposed for the free nucleon 
system~\cite{Goeke, Alkofer} and compare the obtained results. 

This paper is organized as follows. In the following section, we present the basic 
formulation of CQSM with vector mesons.
Two distinct formulations for solving the non-Hermitian eigenvalue problem are reviewed 
in Sec.\ref{sec:level3}.
In Sec.\ref{sec:level4}, we show how various cutoff parameters and coupling constants are 
determined within the chiral perturbation regime.
In Sec.\ref{sec:level5}, the extension of the model to the nuclear matter within the Wigner-Seitz approximation 
is presented. The numerical results are shown in Sec.\ref{sec:level6}. 
Sec.\ref{sec:level7} is devoted to summary and conclusions.

\section{\label{sec:level2}The chiral quark soliton model with ${\rm \rho,\omega}$ mesons \protect\\ }
The CQSM was originally derived from the instanton 
liquid model of the QCD vacuum and incorporates the non-perturbative 
feature of the low-energy QCD, spontaneous chiral symmetry breaking (SCSB). 
The semibosonized version of the Nambu-Jona-Lasinio model also inspires the CQSM model 
with the SCSB. In these description, the Euclidean vacuum functional 
with vector mesons can be defined as~\cite{Goeke,Alkofer}
\begin{eqnarray}
	{\cal Z} &=& \int {\cal D}\pi{\cal D}V{\cal D}\psi{\cal D}\psi^{\dagger}\nonumber\\
	&\times& \exp \left[ \int d^{4}x \, \bar{\psi}\left(i\!\!\not\!\partial+\not\!V
	- MU^{\gamma_{5}}\right) \psi \right]	 \label{vacuum_functional}
\end{eqnarray} 
where $V_\mu=\sum^3_{a=0}V_\mu^a\tau^a/2$ ($\tau^0=\bm{1}$) 
are vector gauge fields for the vector mesons. 
The SU(2) matrix
\begin{eqnarray}
	U^{\gamma_{5}}= \frac{1+\gamma_{5}}{2} U + \frac{1-\gamma_{5}}{2} U^{\dagger} 
\end{eqnarray}
with
\begin{eqnarray}
	U=\exp \left( i \bm{\tau} \cdot \bm{\phi}/f_{\pi} \right)
	=\frac{1}{f_\pi}(\sigma+i\bm{\tau}\cdot\bm{\pi})\,,
\end{eqnarray}
describes the chiral fields. $\sigma$ and $\pi$ represent scalar sigma meson and  
pseudoscalar pion fields respectively. $\psi$ denotes quark fields and $M$ is the 
dynamical quark mass. $f_{\pi} $ is the pion decay constant and experimentally 
$f_{\pi} \sim 93 {\rm MeV}$. 
Since our concern is the tree-level pions and one-loop quarks according 
to the Hartree mean field approach, the kinetic term of the pion fields which 
gives a contribution to higher loops can be neglected. 
Due to the interaction between the valence quarks and the Dirac sea, 
soliton solutions appear as bound states of quarks in the background of self-consistent 
mean chiral field. $N_{c}$ valence quarks fill the each bound state to form a baryon. 
The baryon number is thus identified with the number of bound states filled by 
the valence quarks \cite{kahana84}. 
The $B=1$ soliton solution with only chiral fields has been studied in detail at classical and 
quantum level in Refs.~\cite{diakonov88,reinhardt88,meissner89,
christov96,alkofer96,wakamatsu91}.  

Integrating over the quark fields in Eq.(\ref{vacuum_functional}),
we can obtain the effective action $S_{\rm eff}$ for the mesons 
\begin{eqnarray}
	S_{{\rm eff}}&=&S_{F}+S_{m},\\
	S_{F}&=&-iN_{c}{\rm lndet}\left(i\!\!\not\!\partial
	+\not\!V- MU^{\gamma_{5}}\right)\label{effective_action1},\\
	S_{m}&=&\int d^4x\left(\frac{1}{4g}{\rm tr}(V_\mu V^\mu)\right),\label{effective_action2}
\end{eqnarray}
where $S_m$ is the term derived from the semibosonized version of NJL action~\cite{Alkofer}.
$S_F$ can be divided into real and imaginary parts:
\begin{eqnarray}
	&&S_F=S_R+S_I,\\
	&&S_R=\frac{1}{2}{\rm Tr}\log(D\hspace{-.55em}/^\dagger D\hspace{-.55em}/),\label{actionr}\\
	&&S_I=\frac{1}{2}{\rm Tr}\log((D\hspace{-.55em}/^\dagger)^{-1} D\hspace{-.55em}/)
\end{eqnarray}
where $iD\hspace{-.55em}/ = i\!\!\not\!\partial+\not\!V- MU^{\gamma_5}$ is the modified 
Dirac operator.
After performing the Wick rotation for the Dirac operator, $i.e.$ $x_0=-ix_4$ and $V_0=-iV_4$, 
we can obtain the one-quark Hamiltonian $H$ with vector fields
\begin{eqnarray}
	i\beta D\hspace{-.55em}/ &=&-\partial_\tau - H,\label{dirac_operator}\\
	H&=&{\bm{\alpha}\!\cdot\!{\bm P}}+iV_4+{\bm{\alpha}\!\cdot\!{\bm V}}+
	\beta MU^{\gamma_{5}}\,\,. \label{hamiltonian}
\end{eqnarray}
Here $\tau$ denotes the Euclidean time.
Note that as $\tau,V_4$ is supposed to be Hermitian in Euclidean space,
$H$ is now non-Hermitian.

To obtain the $B=1$ soliton solution, let us impose the hedgehog ansatz on the chiral field 
\begin{eqnarray}
U(\bm{r})&=&\exp(i F(r) \hat{\bm{r}}\cdot \bm{\tau})\nonumber\\
&=&\cos F(r)+i\hat{\bm{r}}\cdot \bm{\tau}\sin F(r).\label{chiral_fields_hedgehog}
\end{eqnarray}
The only possible ansatz for the isoscalar-vector field $\omega_{\mu}$ 
realizing zero grandspin is the one whose spatial components vanish ($\omega_i=0$), 
\begin{eqnarray}
V_\mu^0&=&\omega_\mu=\omega(r)\delta_{\mu4}.
\end{eqnarray}
Parity invariance requires the isoscalar-axialvector meson field $V_4$ to vanish 
in the static limit. Note that $V_\mu^0$ corresponds to the physical $\omega$ meson.
For the isovector and vector meson fields let us impose the spherically symmetric ansatz
\begin{eqnarray}
V_\mu&=&-\frac{1}{2}i\rho^a_\mu \tau^a\,,\nonumber\\
\rho^a_0&=&0,\,\,\rho^a_i=-\epsilon^{aik}\hat{r}^k G(r) 
\end{eqnarray}
where the indices $a$,$i$ and $k$ run from 1 to 3.
$V_i^a$ corresponds to the physical $\rho$ meson.
The boundary conditions of $F(r)$ for the $B=1$ soliton solution are given by 
\begin{eqnarray}
F(0)&=&-\pi\,,~~F(\infty)=0\,. \label{boundary_condition_f}
\end{eqnarray}
Regularity requires the following boundary conditions for $\omega$ and G,  
\begin{eqnarray}
\omega^\prime(0)&=&0,\,\,\omega(\infty)=0\nonumber\,,\\
G(0)&=&0,\,\,G(\infty)=0.\label{boundary_condition_wg}
\end{eqnarray}
Substituting these ansatz into Eq.(\ref{hamiltonian}), 
one obtains the effective Hamiltonian
\begin{eqnarray}
H=\mbox{\boldmath $ \alpha\cdot p$}+i\omega(r)
+\frac{1}{2}(\mbox{\boldmath $ \alpha$}\times\hat{{\bm r}})\cdot\mbox{\boldmath $\tau$}G(r)\nonumber\\
+\beta M(\cos{F(r)}+i\gamma_5\mbox{\boldmath $ \tau\cdot \hat{r}$}\sin{F(r)}).
\label{hamiltonian_hedgehog}
\end{eqnarray}
As stated above, $H$ is non-Hermitian since the real function of $\omega(r)$ makes 
$H$ complex-valued.  
The eigenvalue problem of $H$ is solved using the method developed by  
Kahana-Ripka (\cite{kahana84}, see also Sec.\ref{sec:level5}).

Once the eigenvalue of $H$, $\epsilon_\mu$, is obtained,
the eigenvalues $\lambda_{n,\mu}$ of the operator $\partial_\tau+H$ (\ref{dirac_operator}) 
are determined by 
\begin{eqnarray}
\lambda_{n,\mu}=-i\Omega_n+\epsilon_\mu=-i\Omega_n+\epsilon_\mu^R+i\epsilon_\mu^I\label{lambda}
\end{eqnarray}
where $i\Omega_n=i(2n+1)\pi/T$ with $(n=0, \pm 1,\pm 2, ..)$ 
and $\epsilon_\mu^R, \epsilon_\mu^I$ are the real and imaginary part of 
eigenvalues of the Hamiltonian (\ref{hamiltonian_hedgehog}).
The quark determinant is expressed in terms of the eigenvalues $\lambda_{n,\mu}$ as 
\begin{eqnarray}
S_R=\frac{1}{2}\sum_{\mu,n}\log(\lambda_{n,\mu}\lambda_{n,\mu}^*),
\ S_I=\frac{1}{2}\sum_{\mu,n}\log\left(\frac{\lambda_{n,\mu}}{\lambda_{n,\mu}^*}\right).
\end{eqnarray}
Since the real part $S_R$ diverges as log $p^2$ for large momenta $p$,
we apply the proper time regularization~\cite{Alkofer}.
The real part of the sea quark energy from can be derived from 
${\rm e}^{S_R}\sim  {\rm e}^{-E_{\rm vac}^RT}$ as  
\begin{eqnarray}
E_{\rm vac}^R=\frac{N_c}{4 \sqrt{\mathstrut \pi}}\sum_\mu |\epsilon_\mu^R|
\Gamma\left(-\frac{1}{2},\left(\frac{\epsilon_\mu^R}{\Lambda}\right)^2\right).
\end{eqnarray}
Similarly the imaginary part of the sea quark energy is derived from 
${\rm e}^{S_I}\sim  {\rm e}^{-iE_{\rm vac}^IT}$ as 
\begin{eqnarray}
E_{\rm vac}^I=N_c\sum_\mu\epsilon_\mu^I{\rm sign}(\epsilon_\mu^R){\cal N}_\mu.
\end{eqnarray}
where
\begin{eqnarray}
{\cal N_\mu}=-\frac{1}{\sqrt{\mathstrut 4\pi}}\Gamma\left(\frac{1}{2},
\left(\frac{\epsilon_\mu^R}{\Lambda}\right)^2\right).
\end{eqnarray}
The static energy for the vector mesons is given by  
\begin{eqnarray}
E_{m}=\frac{1}{2}\frac{1}{4g_1}\int d^3xG(r)^2+\frac{1}{2}\frac{1}{4g_2}
\int d^3x\omega(r)^2\label{meson_term}\,.
\end{eqnarray}
where $g_1$ and $g_2$ are coupling constants determined in the 
subsequent section.
The total energy $E_{\rm tot}$ is defined by the sum of these energies plus 
(three times of) valence quark energy (see Sec.\ref{sec:level3}). 
Field equations for the meson fields can be obtained by demanding  
that the total energy be stationary 
with respect to variation of the profile function,
\begin{eqnarray}
\frac{\delta E_{\rm tot}}{\delta \phi}=0 \label{variation_profile}
\end{eqnarray}
where $\phi$ denotes any of the meson profile $F,G$ or $\omega$, 
which produces 
\begin{eqnarray}
S(r)\sin{F(r)}=P(r)\cos{F(r)},\nonumber
\end{eqnarray}
\begin{eqnarray}
S(r)&=&N_c{\rm tr}\int\frac{d\Omega}{4\pi}\gamma_0
\rho(\mbox{\boldmath $x$},\mbox{\boldmath $x$}),\nonumber\\
P(r)&=&N_c{\rm tr}\int\frac{d\Omega}{4\pi}
\left(i\gamma_0\gamma_5\mbox{\boldmath $\hat{r}$}\cdot\mbox{\boldmath $\tau$})\right)
\rho(\mbox{\boldmath $x$},\mbox{\boldmath $x$})\nonumber\\
G(r)&=& - g_1 N_c{\rm tr}\int\frac{d\Omega}{4\pi}
\left((\mbox{\boldmath $\gamma$}\times\mbox{\boldmath $\hat{r}$})\cdot\mbox{\boldmath $\tau$}\right)
\rho(\mbox{\boldmath $x$},\mbox{\boldmath $x$})\nonumber\\
w(r)&=& g_2 N_c{\rm tr}\int\frac{d\Omega}{4\pi}
b(\mbox{\boldmath $x$},\mbox{\boldmath $x$})\label{eq}
\end{eqnarray}
where $\rho(\mbox{\boldmath $x$},\mbox{\boldmath $y$})$, 
 $b(\mbox{\boldmath $x$},\mbox{\boldmath $y$})$ 
are the quark scalar density and the quark number density respectively.
Their specific forms are presented in the next section.

\section{\label{sec:level3}The eigenvalue problem for the non-Hermitian Hamiltonian }

In order to solve the eigenvalue problem with the non-Hermitian Hamiltonian, let us 
first introduce the left and right eigenstate
\begin{eqnarray}
H\mid\hspace{-1mm}\Psi_\mu\rangle&=&\epsilon_\mu\mid\hspace{-1mm}\Psi_\mu\rangle~,\nonumber\\
\langle\tilde{\Psi}_\mu\hspace{-1mm}\mid H&=&\epsilon_\mu\langle\tilde{\Psi}_\mu\hspace{-1mm}\mid\,,
\ \ \ i.e.\ \ H^\dagger\mid\hspace{-1mm}\tilde{\Psi}_\mu\rangle=\epsilon_\mu^*\mid\hspace{-1mm}
\tilde{\Psi}_\mu\rangle
\label{lreigen}
\end{eqnarray}
with the normalization condition $\langle\tilde{\Psi}_\nu\hspace{-1mm}\mid\hspace{-1mm}
\Psi_\mu\rangle=\delta_{\mu\nu}$.
For convenience we separate the Hamiltonian into Hermitian and non-Hermitian part  
\begin{eqnarray}
H=H_\Theta+i\omega\,
\end{eqnarray}
where $H_\Theta$ is the Hermitian part.
The fact that $H_\Theta$ and $\omega$ are both Hermitian implies
$\mid\hspace{-1mm}\tilde{\Psi}_\mu\rangle=\mid\hspace{-1mm}\Psi_\mu^*\rangle$.
There are two distinct ways \cite{Goeke,Alkofer} 
to extract the physical spectra from the eigenequations (\ref{lreigen}).

\begin{figure}
\includegraphics[height=7cm, width=9cm]{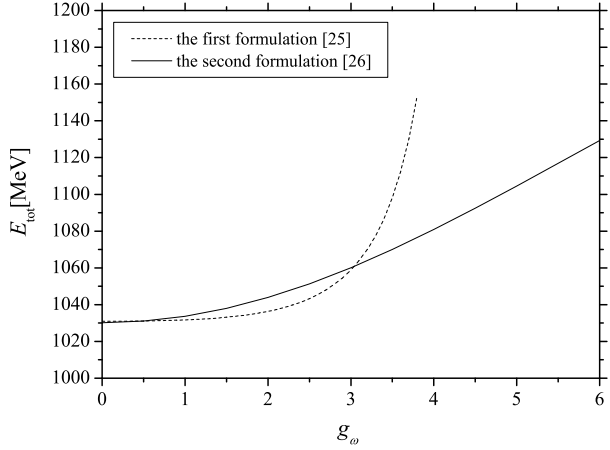}
\caption{\label{fig:Fig1} The total energy as a function of $g_\omega$ 
in the formalisms of Refs.\cite{Goeke} and \cite{Alkofer}. 
In the formalism of Ref.\cite{Goeke}, the solution does not exist for $g_\omega \gtrsim  3.8$. 
$g_\omega\sim 4.6$ is predicted in the chiral perturbation analysis (see Sec.5).}
\end{figure}

In Ref.\cite{Goeke}, the Wick rotation from Euclidean to Minkowski space has been 
performed to the Hamiltonian.
Since the time component of the vector fields becomes $\omega_4\to i\omega_0$
the eigen equations are reduced to
\begin{eqnarray}
H\mid\hspace{-1mm}\Psi_\mu\rangle=\epsilon_\mu\mid\hspace{-1mm}
\Psi_\mu\rangle\to H^{(+)}\mid\hspace{-1mm}\Psi_\mu\rangle^{(+)}
=\epsilon_\mu^+\mid\hspace{-1mm}\Psi_\mu\rangle^{(+)},\nonumber\\
H^\dagger\mid\hspace{-1mm}\Psi_\mu\rangle=\epsilon_\mu^*\mid
\hspace{-1mm}\Psi_\mu\rangle\to H^{(-)}\mid\hspace{-1mm}
\Psi_\mu\rangle^{(-)}=\epsilon_\mu^-\mid\hspace{-1mm}\Psi_\mu\rangle^{(-)}.
\end{eqnarray}
Defining   
\begin{eqnarray}
	\epsilon_\mu^R=\frac{1}{2}(\epsilon_\mu^++\epsilon_\mu^-),\,\,
	\epsilon_\mu^I=\frac{1}{2}(\epsilon_\mu^+-\epsilon_\mu^-), \label{}
\end{eqnarray} 
one can write the valence quark energy as 
\begin{eqnarray}
E_{\rm val}=N_cn_{\nu}\epsilon^R_{\rm val}
\end{eqnarray}
where $n_\nu$ is the valence quark occupation number.
Then the total energy is given by  
\begin{eqnarray}
E_{\rm tot}[F,\omega,G]=E_{\rm val}+E_{\rm vac}^R+E_{\rm vac}^I+E_m. \label{total_geoke}
\end{eqnarray}
By substituting (\ref{total_geoke}) into Eq.(\ref{variation_profile}),
we obtain the equation of motion (\ref{eq}) where the quark scalar density and 
the quark number density are given by
\begin{eqnarray}
\rho(\mbox{\boldmath $x$},\mbox{\boldmath $y$})
&=&F^+\bar{\psi}\psi^++F^-\bar{\psi}^-\psi^-\nonumber\\
b(\mbox{\boldmath $x$},\mbox{\boldmath $y$})
&=&F^+\bar{\psi}\psi^+-F^-\bar{\psi}^-\psi^- ,
\end{eqnarray}
\begin{eqnarray}
	F^+&=&\left(n_\mu-\frac{1}{4}\sum_\mu{\rm sign}(\epsilon_\mu^R)
	(1-2{\mathcal N}_\mu)\right)  \label{},\nonumber\\
	F^-&=&\left(\frac{1}{4}\sum_\mu{\rm sign}(\epsilon_\mu^R)
	(1+2{\mathcal N}_\mu)\right) \label{}.
\end{eqnarray}
Unfortunately, this simple method do not produce solutions for the predicted value of the $\omega$ 
meson coupling constant $g_\omega$ required in the chiral perturbation analysis. 
In Fig. \ref{fig:Fig1}, we show the total energy of the soliton as a function of $g_\omega$.
As can be seen, the soliton survives only up to $g_\omega\sim3.8$ whereas 
the chiral perturbation analysis predicts the value around $g_\omega\sim 4.6$.

\begin{figure}
\includegraphics[height=7cm, width=9cm]{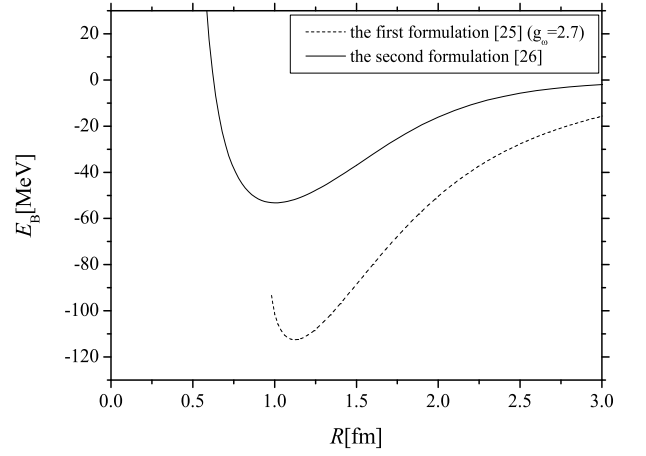}
\caption{\label{fig:Fig2} The binding energy in the formalisms of Refs.\cite{Goeke} and \cite{Alkofer},
where $g_\omega$=2.7 is used for the former and $g_\omega\sim4.6$ for the latter.}
\end{figure}

In the second method, Eq.~(\ref{lreigen}) is solved directly. 
Then the real and imaginary part of the one particle energy eigenvalue are derived as \cite{Alkofer}.
\begin{eqnarray}
\epsilon_\mu^R&=&\frac{1}{2}(\langle\Psi_\mu^*\mid H\mid\hspace{-1mm}\Psi_\mu\rangle
+\langle\Psi_\mu\mid H\mid\hspace{-1mm}\Psi_\mu^*\rangle)\nonumber\\
&=&\langle\Psi_\mu^R\mid H_\Theta\mid\hspace{-1mm}\Psi_\mu^R\rangle-\langle\Psi_\mu^I\mid 
H_\Theta\mid\hspace{-1mm}\Psi_\mu^I\rangle\nonumber\\
&-&\langle\Psi_\mu^I\mid \omega\mid\hspace{-1mm}
\Psi_\mu^R\rangle-\langle\Psi_\mu^R\mid \omega\mid\hspace{-1mm}\Psi_\mu^I\rangle,\nonumber\\
\epsilon_\mu^I&=&\frac{1}{2}(\langle\Psi_\mu^*\mid H\mid\hspace{-1mm}\Psi_\mu\rangle
-\langle\Psi_\mu\mid H\mid\hspace{-1mm}\Psi_\mu^*\rangle)\nonumber\\
&=&\langle\Psi_\mu^R\mid \omega\mid\hspace{-1mm}\Psi_\mu^R\rangle-\langle\Psi_\mu^I\mid 
\omega\mid\hspace{-1mm}\Psi_\mu^I\rangle\nonumber\\
&+&\langle\Psi_\mu^I\mid H_\Theta\mid\hspace{-1mm}
\Psi_\mu^R\rangle+\langle\Psi_\mu^R\mid H_\Theta\mid\hspace{-1mm}\Psi_\mu^I\rangle
\end{eqnarray}
where we employed the decomposition 
$\mid\hspace{-1mm}\Psi_\mu\rangle=\mid\hspace{-1mm}\Psi_\mu^R\rangle+i\mid\hspace{-1mm}\Psi_\mu^I\rangle$ 
and $\langle\Psi_\mu^*\mid=\langle\Psi_\mu^R\mid+i\langle\Psi_\mu^I\mid$.
The valence quark energy is given by the same form as in the first method 
\begin{eqnarray}
	E_{\rm val}^{R}=N_c\sum_\nu n_\nu |\epsilon_\nu^R|,\, 
	E_{\rm val}^{I}=N_c\sum_\nu n_\nu {\rm sign}(\epsilon_\nu^R)|\epsilon_\nu^I| \label{}.
\end{eqnarray}
The total energy in Euclidean space is
\begin{eqnarray}
E_{\rm tot}[F,\omega,G]=E_{\rm val}^R+E_{\rm vac}^R+i(E_{\rm val}^I+E_{\rm vac}^I)+E_m.
\end{eqnarray}
Thereby the total energy in Minkowski space is interpreted as 
\begin{eqnarray}
E_{\rm tot}[F,\omega,G]=E_{\rm val}^R+E_{\rm val}^I+E_{\rm vac}^R+E_{\rm vac}^I+E_m.
\end{eqnarray}
The equation of motion takes the form in (\ref{eq}) with the replacement of $\rho(\mbox{\boldmath $x$},
\mbox{\boldmath $y$})$ and $\rho(\mbox{\boldmath $x$},\mbox{\boldmath $y$})$ by
\begin{eqnarray}
\rho(\mbox{\boldmath $x$},\mbox{\boldmath $y$})=
\rho_R^{\rm val}+\rho_I^{\rm val}+\rho_R^{\rm vac}+\rho_i^{\rm vac},
\end{eqnarray}
\begin{eqnarray}
&&\rho_R^{\rm val}(\mbox{\boldmath $x$},\mbox{\boldmath $y$})
+\rho_R^{\rm vac}(\mbox{\boldmath $x$},\mbox{\boldmath $y$})\nonumber\\
&&~~=\sum_\nu \left[\psi_\nu^R(\mbox{\boldmath $x$})\bar{\psi}_\nu^R(\mbox{\boldmath $y$})
-\psi_\nu^I(\mbox{\boldmath $x$})\bar{\psi}_\nu^I(\mbox{\boldmath $y$})\right](n_\nu +f_\nu^R),\nonumber\\
&&\rho_I^{\rm val}(\mbox{\boldmath $x$},\mbox{\boldmath $y$})
+\rho_I^{\rm vac}(\mbox{\boldmath $x$},\mbox{\boldmath $y$})\nonumber\\
&&~~=\sum_\nu \left[\psi_\nu^R(\mbox{\boldmath $x$})\bar{\psi}_\nu^I(\mbox{\boldmath $y$})
+\psi_\nu^I(\mbox{\boldmath $x$})\bar{\psi}_\nu^R(\mbox{\boldmath $y$})\right](n_\nu +f_\nu^I),\nonumber \\
	 \label{}
\end{eqnarray}
\begin{eqnarray}
b(\mbox{\boldmath $x$},\mbox{\boldmath $y$})=b_R^{\rm val}+b_I^{\rm val}+b_R^{\rm vac}+b_i^{\rm vac},
\end{eqnarray}
\begin{eqnarray}
&&b_R^{\rm val}(\mbox{\boldmath $x$},\mbox{\boldmath $y$})
+b_R^{\rm vac}(\mbox{\boldmath $x$},\mbox{\boldmath $y$})\nonumber\\
&&~~=\sum_\nu \left[\psi_\nu^R(\mbox{\boldmath $x$})\psi_\nu^{R\dagger}(\mbox{\boldmath $y$})
-\psi_\nu^I(\mbox{\boldmath $x$})\psi_\nu^{I\dagger}(\mbox{\boldmath $y$})\right](n_\nu +f_\nu^I),\nonumber\\
&&b_I^{\rm val}(\mbox{\boldmath $x$},\mbox{\boldmath $y$})
+b_I^{\rm vac}(\mbox{\boldmath $x$},\mbox{\boldmath $y$})\nonumber\\
&&~~= - \sum_\nu \left[\psi_\nu^R(\mbox{\boldmath $x$})\psi_\nu^{I\dagger}(\mbox{\boldmath $y$})
+\psi_\nu^I(\mbox{\boldmath $x$})\bar{\psi}_\nu^{R\dagger}(\mbox{\boldmath $y$})\right](n_\nu +f_\nu^R),\nonumber\\
	 \label{}
\end{eqnarray}
where $f_R$ and $f_I$ are the imaginary part of the reguralized action written by
\begin{eqnarray*}
	f_\nu^R&=&{\rm sign}(\epsilon_\nu^R){\mathcal N}_\nu+\frac{1}{\sqrt{\pi}}(\epsilon_\nu^I/\Lambda)
	\exp(-(\epsilon_\nu^R/\Lambda)^2),\nonumber\\
	f_\nu^I&=&{\rm sign}(\epsilon_\nu^R){\mathcal N}_\nu .
\end{eqnarray*}
 
\begin{figure}
\includegraphics[height=7cm, width=9cm]{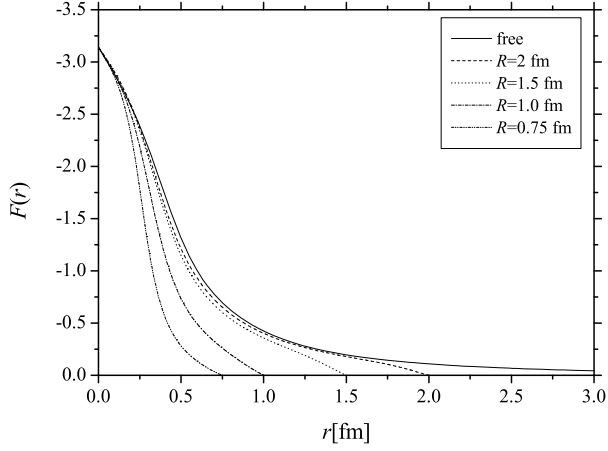}
\caption{\label{fig:Fig3} The profile functions of the chiral fields 
for $R$=0.75,1,1.5,2fm and the free ($R\to\infty$) solution.}
\end{figure}

\section{\label{sec:level4}The coupling constants of the vector mesons \protect}
In this section, we derive the coupling constants for the vector mesons. 
Expanding the real part of the effective action~(\ref{actionr}) up to second order 
in the heat-kernel expansion~\cite{ebert86}, one obtains 
\begin{eqnarray}
&&S_f^{R(2)}=\frac{N_c}{16\pi^2}\int d^4x\Gamma \left(0,\left(\frac{M}{\Lambda}\right)^2\right)\nonumber\\
&&\times{\rm tr}\left(M^2\partial_\mu U\partial^\mu U^\dagger 
+\frac{1}{3}F_{\mu \nu}^k F^{\mu \nu}_k
+\frac{1}{3}F_{\mu \nu}^0 F^{\mu \nu}_0\right),~~\label{expand_s}
\end{eqnarray}
where
\begin{eqnarray}
F_{\mu \nu}^k&=&\partial_\mu V^k_\nu-\partial_\nu V^k_\mu +[V_\mu^k , V_\nu^k],\nonumber\\ 
F_{\mu \nu}^0&=&\partial_\mu V^0_\nu-\partial_\nu V^0_\mu +[V_\mu^0 , V_\nu^0].\nonumber
 \label{}
\end{eqnarray}
We renormarize the vector meson fields
$V_\mu^k =-ig_\rho \tilde{V}_\mu^k,\ V_\mu^0 =-ig_\omega \tilde{V}_\mu^0$
for later convenience.
Then the total Lagrangian is given by
\begin{eqnarray}
L^{R(2)}&=&\frac{N_c}{16\pi^2}\Gamma \left(0,\left(\frac{M}{\Lambda}\right)^2\right)
{\rm tr} \biggl[ M^2\partial_\mu U\partial^\mu U^\dagger \nonumber\\
&+&\frac{g_\rho^2}{3}F_{\mu \nu}^k F^{\mu \nu}_k
+\frac{g_\omega^2}{3}F_{\mu \nu}^0 F^{\mu \nu}_0\biggl] \nonumber\\
&+&\frac{1}{2}\left(\frac{g_\rho^2}{4g_1}\right)(\tilde{V}^k)^2
+\frac{1}{2}\left(\frac{g_\omega^2}{4g_2}\right)(\tilde{V}^0)^2 \label{expand_lag}
\end{eqnarray}
where
\begin{eqnarray}
\tilde{F}_{\mu \nu}^k&=&\partial_\mu \tilde{V}^k_\nu-\partial_\nu \tilde{V}^k_\mu
 +g_\rho[\tilde{V}_\mu^k , \tilde{V}_\nu^k],\nonumber\\ 
\tilde{F}_{\mu \nu}^0&=&\partial_\mu \tilde{V}^0_\nu-\partial_\nu \tilde{V}^0_\mu 
+g_\omega[\tilde{V}_\mu^0 , \tilde{V}_\nu^0] \label{}.
\end{eqnarray}
Comparing Eq.~(\ref{expand_lag}) with the massive Yang-Mills Lagrangian,
the following relations for the parameters are obtained, 
\begin{eqnarray}
f_\pi^2&=&\frac{N_c M^2}{4\pi^2}\Gamma \left(0,\left(\frac{M}{\Lambda}\right)^2\right),\\
g_\rho^2&=&\frac{6M^2}{f_\pi^2},\ g_\omega^2=\frac{6M^2}{4f_\pi^2},\\
M_\rho^2&=&\frac{g_\rho^2}{4g_1},\ M_\omega^2=\frac{g_\omega^2}{4g_2}. \label{}
\end{eqnarray}
The experimental values are  
$f_\pi$=93MeV, $M$=350MeV, $M_\rho$=770MeV, $M_\omega$=783MeV, 
and hence 
\begin{eqnarray}
&&\Lambda=649{\rm MeV},~g_1=1.39,~g_2=0.34,\nonumber \\
&&\hspace{1cm}g_\rho=9.22,~g_\omega=4.61. 
\end{eqnarray}
As a reference, let us note that in Ref.~\cite{saito94}, 
\begin{eqnarray}
g_\rho=8.32\,,~g_\omega=4.41 
\end{eqnarray}
are adopted for the MIT bag model.   

\begin{figure}[t]
\includegraphics[height=7cm, width=9cm]{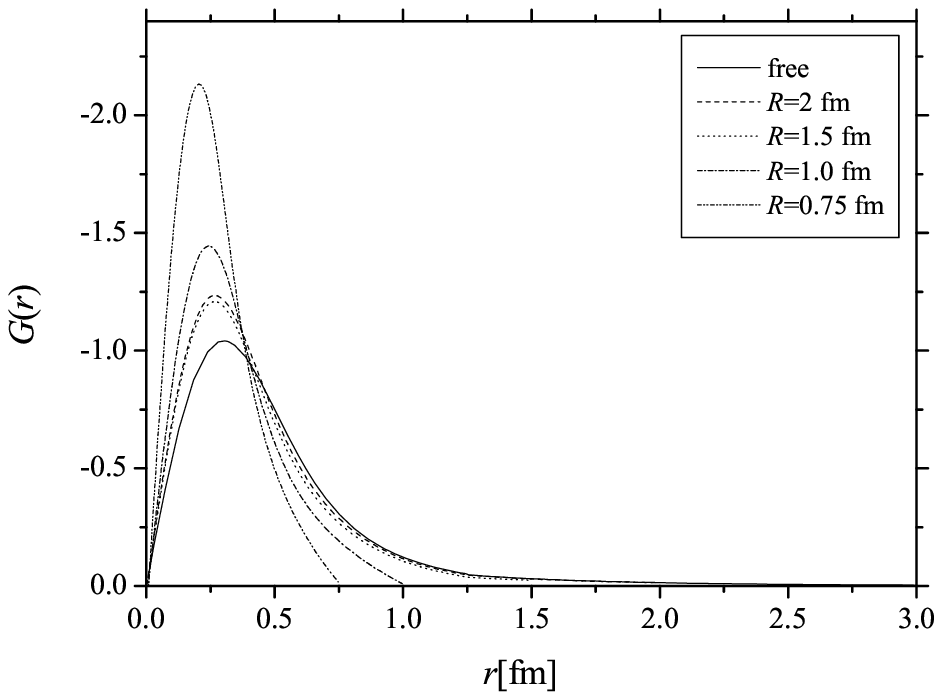}
\caption{\label{fig:Fig4} The profile functions of $\rho$ meson 
for $R$=0.75,1,1.5,2fm and the free ($R\to\infty$) solution.}
\end{figure}

\section{\label{sec:level5}The nuclear matter formulation}
In this section we describe the numerical method of eigenproblem of the 
Hamiltonian~(\ref{hamiltonian_hedgehog}). The Hamiltonian with hedgehog 
ansatz commutes with the parity and the grandspin operator given by  
\begin{eqnarray*}
	\bm{K}=\bm{j}+\bm{\tau}/2=\bm{l}+\bm{\sigma}/2+\bm{\tau}/2,
\end{eqnarray*}
where $\bm{j},\bm{l}$ are respectively total angular momentum and orbital angular momentum. 
Accordingly, the angular basis can be written as  
\begin{eqnarray}
|(lj)KM\rangle= \sum_{j_3\tau_3}C^{KM}_{jj_3\frac{1}{2}\tau_3}
\Bigl(\sum_{m\sigma_3}C^{jj_3}_{lm\frac{1}{2}\sigma_3}
|lm \rangle |\frac{1}{2}\sigma_3 \rangle \Bigr) |\frac{1}{2} \tau_3 \rangle\,.\nonumber\\
\end{eqnarray}
For $B=1$ solution, following states are possible 
\begin{eqnarray}
&&|0\rangle =|(K~K+\frac{1}{2})KM \rangle\,,  \nonumber   \\
&&|1\rangle =|(K~K-\frac{1}{2})KM \rangle\,,  \nonumber   \\
&&|2\rangle =|(K+1 K+\frac{1}{2})KM\rangle\,, \nonumber  \\
&&|3\rangle =|(K-1 K-\frac{1}{2})KM\rangle\,. \nonumber
\end{eqnarray}
With this angular basis, the normalized eigenstates of the free Hamiltonian 
in a spherical box with radius $R$ can be constructed as follows:
\begin{eqnarray}
&&u^{(a)}_{KM}=
N_k\left( 
\begin{array}{c}
i\omega^{+}_{\epsilon_k}j_{K}(kr)|0\rangle \\
\omega^{-}_{\epsilon_k}j_{K+1}(kr)|2\rangle
\end{array}
\right), \nonumber \\
&&u^{(b)}_{KM}=
N_k\left( 
\begin{array}{c}
i\omega^{+}_{\epsilon_k}j_{K}(kr)|1\rangle \\
-\omega^{-}_{\epsilon_k}j_{K-1}(kr)|3\rangle
\end{array}
\right), \nonumber \\
&&v^{(a)}_{KM}=
N_k\left( 
\begin{array}{c}
i\omega^{+}_{\epsilon_k}j_{K+1}(kr)|2\rangle \\
-\omega^{-}_{\epsilon_k} j_{K}(kr)|0\rangle
\end{array}
\right), \nonumber \\
&&v^{(b)}_{KM}=
N_k\left( 
\begin{array}{c}
i\omega^{+}_{\epsilon_k}j_{K-1}(kr)|3\rangle \\
\omega^{-}_{\epsilon_k} j_{K}(kr)|1\rangle
\end{array}
\right), 
\label{kahana_ripka}
\end{eqnarray}
with
\begin{eqnarray}
	N_k=\biggl[\frac{1}{2}R^3
	\Bigl(j_{K+1}(kR)\Bigr)^2\biggr]^{-1/2}
\end{eqnarray}
and $\omega^{+}_{\epsilon_k>0},\omega^{-}_{\epsilon_k<0}={\rm sgn}(\epsilon_k), 
\omega^{-}_{\epsilon_k>0},\omega^{+}_{\epsilon_k<0}=k/(\epsilon_k+M)$.
The  $u$ and $v$ correspond to the {\it ``natural''} 
and {\it ``unnatural''} components of the basis  
which stand for parity $(-1)^{K}$ and $(-1)^{K+1}$ respectively. 
The momenta are discretized by the boundary conditions $j_K(k_i R)=0$. 
The orthogonality of the basis is then satisfied by  
\begin{eqnarray}
&&\int^R_0 dr r^2 j_K(k_i r)j_K(k_j r) \nonumber \\
&&=\int^R_0 dr r^2 j_{K\pm 1}(k_i r)j_{K\pm 1}(k_j r)  \nonumber \\
&&=\delta_{ij}\frac{R^3}{2}  [j_{K\pm 1}(k_i R)]^2 \, .
\label{orthogonality}
\end{eqnarray}

\begin{figure}[t]
\includegraphics[height=7cm, width=9cm]{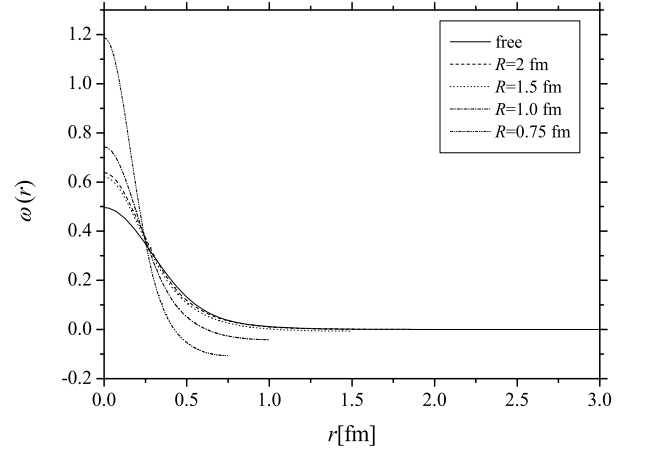}
\caption{\label{fig:Fig5} The profile functions of $\omega$ meson 
for $R$=0.75,1,1.5,2fm and the free ($R\to\infty$) solution.}
\end{figure}

Let us examine the boundary conditions for the Dirac and chiral fields 
to construct the nuclear matter solution in the Wigner-Seitz approximation.
When the background chiral fields are periodic with lattice vector $\bm{a}$, 
the quark fields would be replaced by Bloch wave functions as  
$\psi(\bm{r}+\bm{a})=e^{i\bm{k}\cdot \bm{a}}\psi(\bm{r})$. 
In the Wigner-Seitz approximation, however, the soliton is put on the 
center of the spherical unit cell with the radius $R$ ($a=2R$) 
and the dispersion $\bm{k}$ is assumed to be zero.
Then, $R$ is related to the baryon density through the relation
\begin{eqnarray}
	 \rho_{\rm B} = \frac{3}{4\pi R^3}.\label{R_density}
\end{eqnarray}
For the Dirac eigenstates, modification in the basis is needed. 
For odd number of $K$, the boundary condition is same as the free case with 
\begin{eqnarray}
j_K(k_i R)=0\,.
\end{eqnarray}
For even $K$, the following conditions must be satisfied 
\begin{eqnarray}
j_{K+1}(k^{(a)}_i R)=0,~~{\rm for}~~ u^{(a)}_{KM},v^{(a)}_{KM}\,,\nonumber \\
j_{K-1}(k^{(b)}_i R)=0,~~{\rm for}~~ u^{(b)}_{KM},v^{(b)}_{KM}\,.
\label{mboundary}
\end{eqnarray}
From the conditions (\ref{mboundary}) together with the equations of motion (\ref{eq}), 
we find the following boundary conditions for the profile function $F(r)$
\begin{eqnarray}
\left.
\begin{array}{c} 
~\sigma'(0)=\sigma'(R)=0\\
\pi(0)=\pi(R)=0
\end{array}
 \right\} \Rightarrow 
F(0)=-\pi, F(R)=0\,, \label{boundary_f}
\end{eqnarray}
which guarantees the periodicity and the unit topological 
charge inside the cell. Also, for vector meson profiles 
$\omega(r), G(r)$ we find the conditions 
instead of Eq.(\ref{boundary_condition_wg})
\begin{eqnarray}
\omega^\prime(0)&=&0,\,\,\omega^\prime(R)=0, \\
G(0)&=&0,\,\, G(R)=0.    \label{boundary_g}
\end{eqnarray}
We solve (\ref{eq}) selfconsistently 
with boundary conditions (\ref{mboundary})-(\ref{boundary_g}) for varying $R$,
from infinity (isolate the soliton) to origin (infinite density matter).

\begin{figure}[t]
\includegraphics[height=7cm, width=9cm]{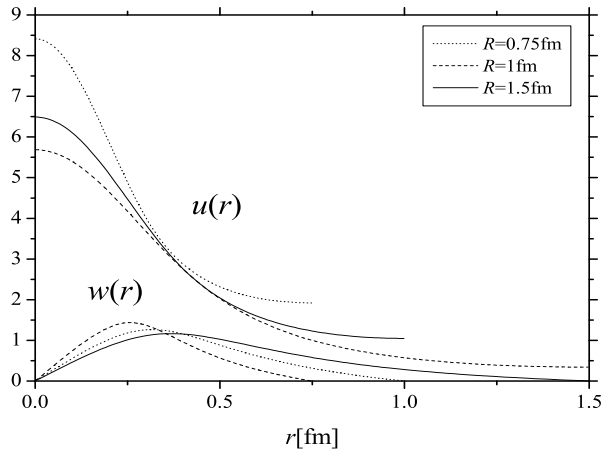}
\caption{\label{fig:Fig6} The ``upper'' $u(r)$ and the ``lower'' $w(r)$ 
positive component of valence quark wave functions for various cell
radius $R$ with the boundary condition $w(R)=0$.}
\end{figure}

\section{\label{sec:level6}The numerical results\protect\\ }
Since the equations of motion for mesons (\ref{eq}) and 
the Dirac equations for quarks (\ref{lreigen}) are highly non-linear, 
we solve these equations numerically by selfconsistent analysis.
Using trial profiles for $\pi, \rho$ and $\omega$ mesons which satisfy  
the appropriate boundary conditions, we solve the Dirac equation (\ref{lreigen}).
From Eq.(\ref{eq}), the profile functions $F(r)$, $G(r)$ and $\omega(r)$ 
are uniquely determined by using the eigenstates of Eq.(\ref{lreigen}).
The new profiles produce new eigenstates.
These procedures are repeated until selfconsistency is attained.
We chose he quark mass $M=350$MeV which is mostly used and turned out to be 
the best choice for obtaining the various experimental observables \cite{christov96,alkofer96}.
We performed the computation from $R$=5 (infinity) upto the value where the soliton 
solution breaks. 
The most dense solution is obtained for $R$=0.5fm which corresponds to the density 
$\rho$=1.91${\rm fm}^{-3}\sim 11\rho_N$. 
Note that $\rho_N$=0.17${\rm fm}^{-3}$ is the standard saturation density.

\begin{figure}[t]
\includegraphics[height=7cm, width=9cm]{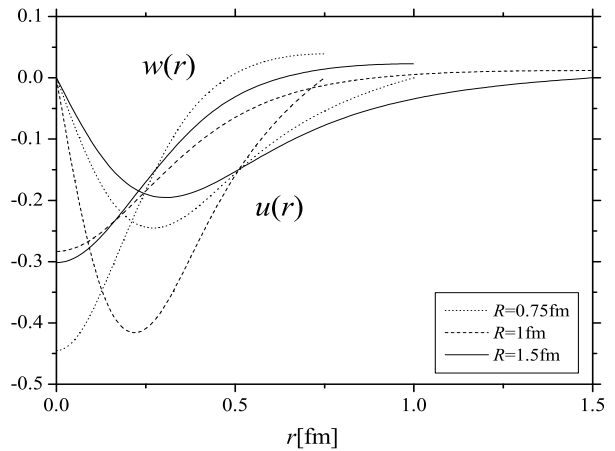}
\caption{\label{fig:Fig7} The negative component of valence quark wave functions.
The label is as same as Fig. \ref{fig:Fig6}.}
\end{figure}

In Fig.~\ref{fig:Fig2}, the binding energies computed in the two methods are shown. 
In the first method, the binding energy is too large, and the solutions disappear at 
relatively lower densities. Therefore, we focus our attention to the numerical results 
obtained in the second formalism hereafter.

Fig.~\ref{fig:Fig3} shows the self-consistent profile functions for free ($R\to \infty$) and 
for the various values of the cell radius $R$. 
Figs.~\ref{fig:Fig4}, \ref{fig:Fig5} show meson profile functions 
of $G(r),\omega(r)$ respectively. Fig. \ref{fig:Fig6} shows the real part of the 
quark wave functions.
Nonvanishing values of the upper component at the cell boundary $u(R)$ 
come from the zero-mode elements in the basis.  
The imaginary part of the quark wave functions derived in the second formulation 
is shown in Fig. \ref{fig:Fig7}.

\begin{figure}[t]
\includegraphics[height=7cm, width=9cm]{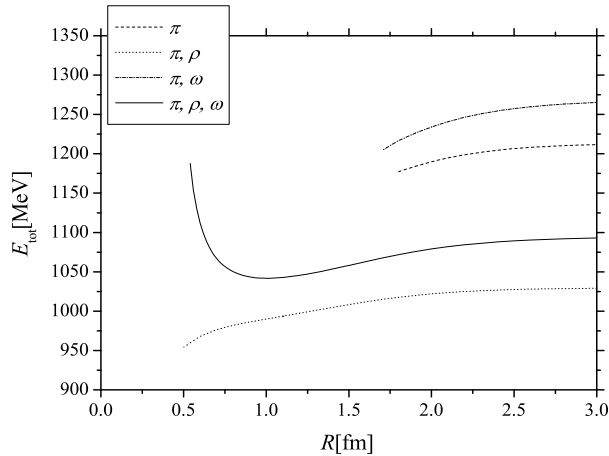}
\caption{\label{fig:Fig8} The total energy of soliton for the various combinations of 
the mesons.}
\end{figure}

In Fig. \ref{fig:Fig8}, we present the results of the 
total energy of the soliton as a function of $R$, for various values  
of the meson couplings. For only the pion coupling, the saturation can not be observed  
and the soliton disappears at very low density.
Adding the $\rho$ meson, the soliton survives at higher density,  
but no saturation is observed.
In the case of $\pi-\omega$, the total energy is enhanced and the soliton disappears 
at low density. The behavior is similar to the case of the pion coupling only.
The saturation can be observed only when all the meson couplings are incorporated.
The saturation point is at $R$=1.0fm corresponding to the density $\rho=0.24{\rm fm}^{-3}$ and 
the binding energy 53MeV.
The density is very close to the experimental value for nuclear matter which is $R$=1.1fm 
corresponding to the density $\rho=0.17{\rm fm}^{-3}$. 
The binding energy is deeper compared to its empirical value, 16MeV.  
In Fig. \ref{fig:Fig9} we show the binding energy for varying the constituent quark mass, 
which is the only free parameter in our model. 
The soliton survives for the constituent quark mass in the range 300MeV $\leq M \leq$ 440MeV.
As $M$ increases, the saturation point moves to the lower density and 
the binding energy becomes shallower.
For $M$=350MeV, the solution reaches to the highest density.

Let us estimate the nuclear incompressibility since it gives an important information for  
the saturation property of the matter.
In Ref. \cite{barnea00}, the authors studied the soliton matter in the Friedberg-Lee model 
with quark-meson coupling using the Wigner-Seitz approximation.
They estimated the incompressibility $K$ with the formula 
\begin{eqnarray}
K=R^2\frac{d^2 E_{\rm B}}{dR^2}
\end{eqnarray}
and obtained $K\sim1170$MeV. The experiment predicts $K$=100-500MeV and generally $K\sim200$MeV.
In our previous analysis \cite{nagai06} with only $\pi$ mesons, we obtained $K\sim400$MeV.
Our new result predicts  $K\sim270$MeV. 

\section{\label{sec:level7}Summary\protect\\ }
In this paper we studied the nuclear matter solutions in the chiral quark soliton model.
We adopted Wigner-Seitz approximation and investigated the saturation property of the matter solutions.
To improve the qualitative behavior at the saturation point, we introduced the $\rho, \omega$ mesons 
in the model. The $\rho$ meson can be incorporated in a straightforward manner. On the other hand, 
Incorporating the $\omega$ meson makes the Hamiltonian complex-valued and requires some technique. 
We tested two distinct formulations proposed in Refs.\cite{Goeke,Alkofer} to solve the non-Hermitian 
eigenvalue problem.
It turned out that the latter method produces the better behavior of solutions especially at 
high-density region. 
We performed the chiral perturbation analysis to determine the meson properties, i.e.,
the coupling constants and the cutoff parameter.
We found that stable nuclear matter solutions exist when $\pi,\rho,\omega$ mesons are included 
with $330\le M \le 420$[MeV].

\begin{figure}[t]
\includegraphics[height=7cm, width=9cm]{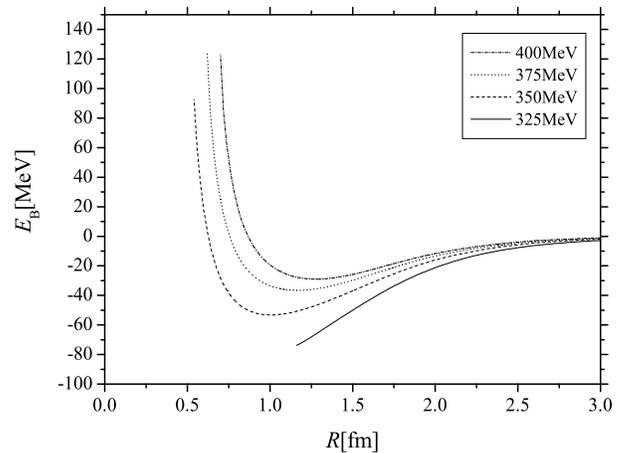}
\caption{\label{fig:Fig9} The total energy varying the constituent quark mass $M$.}
\end{figure}

From Fig. \ref{fig:Fig8}, one can speculate that the attractive 
property of the $\rho$ meson makes the total energy smaller and allows the solution 
to survive at the high density regime. The repulsive property of the $\omega$ meson 
makes the total energy larger and creates a short range (high density) core. 
Thus, the model having both effects can produce a stable nuclear matter. 

\begin{table}[t]
\caption{\label{value_varing_M}We present the $\rho$, $\omega$ coupling constant $g_\rho$, $g_\omega$, 
the total energy of the free nucleon $E_{\rm tot}^{\rm free}$ (its the experimental value is 939MeV), 
the binding energy  $E_{\rm B}$ (16MeV), the saturation density $\rho_s$ (0.17${\rm fm}^{-3}$) 
and the incompressibility $K$ (210$\pm$30MeV)
for the constituent quark mass $M$=350, 375, 400MeV.}
\begin{ruledtabular}
\begin{tabular}{ccccccc}
   $M[{\rm MeV}]$ & $g_\rho$ & $g_\omega$ & $E_{\rm tot}^{\rm free}[{\rm MeV}]$ 
& $E_{\rm B}[{\rm MeV}]$ & $\rho_s[{\rm fm}^{-3}]$ & $K$[MeV] \\ \hline 
350 & 9.22 & 4.61 & 1095 & 53 & 0.24 & 271 \\
375 & 9.88 & 4.94 & 1070 & 36 & 0.13 & 242 \\
400 & 10.54 & 5.27 & 1044 & 29 & 0.11 & 274 \\
\end{tabular}
\end{ruledtabular}
\end{table}

Although our formulation realizes a improved nuclear saturation property, the binding energy 
is still large.
Throughout the calculation, we set the constituent quark mass $M$=350MeV, in which the
solution survives at highest density. 
By varying the value of $M$ in a few ten MeV, the saturation property should be 
improved (see Fig. \ref{fig:Fig9}).
TABLE \ref{value_varing_M} shows the computed saturation properties for $M$=350, 375, 400MeV.
$M$=375MeV seems to be the best choice, but the binding energy is
still large. Another attempt to improve the binding energy may be to introduce heavier mesons, 
like axial vector meson $a_1$ ($\pi, \sigma, \rho, \omega$ and $a_1$ are the mesons which do not 
vanish for hedgehogs \cite{Goeke}).
It is known that $a_1$ meson has attractive property like $\rho$ meson and then it would make the 
matter softer. Taking account the Fermi motion would also make the saturation energy shallower.

\begin{center}
	{\bf Acknowledgements}
\end{center}
We are grateful to Kouichi Saito for fruitful discussions and comments.


\begin{thebibliography}{qq} \bibitem{klebanov85}
Igor Klebanov, Nucl. Phys. B {\bf 262}, 133 (1985). \bibitem{wust87} E. W\"ust, B. E. Brown and A. D. Jackson, 
Nucl. Phys. A {\bf 468}, 450 (1987). \bibitem{manton87} Alfred S. Goldhaber and N. S. Manton, 
Phys. Lett. B {\bf 19}, 231 (1987). \bibitem{castillejo89} L. Castillejo, P. S. Jones, A. D. Jackson, 
J. J. M. Verbaarschot and A. Jackson, 
Nucl. Phys. A {\bf 501}, 450 (1987).
\bibitem{kugler89}
M. Kugler and S. Shtrikman, 
Phys. Rev. D {\bf 40}, 3421 (1989). \bibitem{park02} Byung-Yoon Park, Dong-Pil Min, Mannque Rho and  Vincente Vento, Nucl. Phys. A {\bf 707}, 381 (2002). \bibitem{achtzehnter85} Joachim Achtzehnter, Werner Scheid and Lawrence Wilets, 
Phys. Rev. D {\bf 32}, 2414 (1985).  \bibitem{banerjee85}
B. Banerjee, N. K. Glendenning and V. Soni, 
Phys. Lett. B {\bf 155}, 213 (1985).
\bibitem{glendenning86}
N. K. Glendenning and B. Banerjee, 
Phys. Rev. C {\bf 34}, 1072 (1986).
\bibitem{hahn87}
Detlev Hahn and Norman K. Glendenning, 
Phys. Rev. C {\bf 36}, 1181 (1987).
\bibitem{weber98}
Urban Weber and Judith A. McGovern, 
Phys. Rev. C {\bf 57}, 3376 (1998). 
\bibitem{reinhardt85}
H. Reinhardt, B. V. Dang, and H. Schulz, 
Phys. Lett. B {\bf 159}, 161 (1985).
\bibitem{birse88}
M. C. Birse, J. J. Rehr and L. Wilets, 
Phys. Rev. C {\bf 38}, 359 (1988).
\bibitem{barnea00}
Nir Barnea, Timothy S. Walhout, 
Nucl. Phys. A {\bf 677}, 367 (2000). \bibitem{Kutschera84} M. Kutschera, C. J. Pethick and D. G. Ravenhall, 
Phys. Rev. Lett. {\bf 53}, 1041 (1984).
\bibitem{diakonov88}
D. I. Diakonov, V. Yu. Petrov, and P. V. Pobylitsa, 
Nucl. Phys. B {\bf 306}, 809 (1988).
\bibitem{reinhardt88}
H. Reinhardt and R. W\"unsch , Phys. Lett. B {\bf 215}, 577 (1988).
\bibitem{meissner89}
Th. Meissner, F. Gr\"ummer, and K. Goeke, 
Phys. Lett. B {\bf 227}, 296 (1989).
\bibitem{wakamatsu91}
M. Wakamatsu and H. Yoshiki, Nucl. Phys. A {\bf 524}, 561 (1991).
\bibitem{christov96}
Chr.\ V.\ Christov, A.\ Blotz, H.-C.Kim, P.\ Pobylitsa, T.\ Watabe, Th.\ Meissner, 
E.\ Ruiz Arriola, K.\ Goeke, Prog.\ Part.\ Nucl.\ Phys.\ {\bf 37}, 91 (1996).
\bibitem{alkofer96}
R.\ Alkofer, H.\ Reinhardt and H.\ Weigel, Phys.\ Rept.\ {\bf 265}, 139 (1996)
\bibitem{amore00}
P. Amore and A. De Pace, 
Phys. Rev. C {\bf 61}, 055201 (2000).
\bibitem{adjali92}
I. Adjali, I. J. Aitchison, and J. A. Zuk, 
Nucl. Phys. A {\bf 537}, 457 (1992).
\bibitem{nagai06}
S.Nagai, N.Sawado and N.Shiiki, 
Phys. Lett. B {\bf 632}, 644 (2006).
\bibitem{Goeke}
C. Sch\"uren, F. D\"oring,  
E.\ Ruiz Arriola, K.\ Goeke, Nucl. Phys. A {\bf 565}, 687 (1993).
\bibitem{Alkofer}
R.\ Alkofer, H.\ Reinhardt and H.\ Weigel, Nucl. Phys. A {\bf 570}, 445 (1994)
\bibitem{kahana84}
S. Kahana and G. Ripka, Nucl. Phys. A {\bf 429}, 462 (1984).
\bibitem{ebert86} D.\ Ebert, H.\ Reinhardt, Nucl. Phys. B {\bf 271}, 188 (1986). 
\bibitem{saito94}
K.Saito and A.W.Thomas, Phys. Lett. B {\bf 327}, 9 (1994).
\end{thebibliography}
\end{document}